\documentclass[aps,prb]{revtex4}
\usepackage{graphicx}% Include figure files
\usepackage{bm}% bold math
\usepackage{nicefrac}
\usepackage{amsmath}
\usepackage{tikz}
\usepackage{epstopdf}
\newcommand{\be}{\begin{equation}}
\newcommand{\ee}{\end{equation}}
\newcommand{\bea}{\begin{eqnarray}}
\newcommand{\eea}{\end{eqnarray}}
\newcommand{\beb}{\begin{eqnarray*}}
\newcommand{\eeb}{\end{eqnarray*}}

\begin{document}
%\preprint{LPTMS-xxx}   

\title{Many-body study of a quantum point contact in the fractional quantum Hall regime at $\nu=5/2$}

\author{Paul Soul\'e$^1$}
\author{Thierry Jolicoeur$^1$}
\author{Philippe Lecheminant$^2$}
\affiliation{$^1$Laboratoire de Physique Th\'eorique et Mod\`eles statistiques,
Universit\'e Paris-Sud, 91405 Orsay, France}
\affiliation{$^2$Laboratoire de Physique Th\'eorique et Mod\'elisation, CNRS UMR 8089,
Universit\'e de Cergy-Pontoise, Site de Saint-Martin,  2 avenue Adolphe Chauvin, 95302 Cergy-Pontoise Cedex, France}

\date{July 14th, 2013}
%%%%%%%%%%%%%%%%%%%%%%%%%%%%%%%%%%%%%%%%%%%%%%%%%%%%%%%%%%%%%%%%%%%%%%%%%%%
\begin{abstract}
We study a quantum point contact in the fractional quantum Hall regime at Landau level filling factors
$\nu =1/3$ and $\nu=5/2$. By using exact diagonalizations in the cylinder geometry we identify the edge modes in the presence
of a parabolic confining potential. By changing the sign of the potential we can access both the tunneling through
the bulk of the fluid and the tunneling between spatially separated droplets.
This geometry is realized in the quantum point contact geometry for two-dimensional electron gases.
In the case of the model Moore-Read Pfaffian state at filling factor $\nu = 5/2$
we identify the conformal towers of many-body eigenstates including the non-Abelian sector.
By a Monte-Carlo technique we compute the various scaling exponents
that characterize the edge modes. In the case of hard-core interactions whose ground states are exact model wavefunction we find
equality of neutral and charged velocities for the Pfaffian state both bosonic and fermionic.
\end{abstract}
%\pacs{71.10.Pm,73.43.Cd}
\maketitle
%%%%%%%%%%%%%%%%%%%%%%%%%%%%%%%%%%%%%%%%%%%%%%%%%%%%%%%%%%%%%%%%%%%%%%%%%%%%

\section{introduction}
Modern semiconductor technology has made it possible to fabricate structures with electronic motion
essentially restricted to two dimensions~\cite{DattaBook}. In these so-called two-dimensional electron gases (2DEG) the motion
along the direction perpendicular to the sample plane is frozen into its quantum-mechanical ground state.
This may be realized in heterostructures and in quantum wells. They differ in the shape of the wavefunction
for the frozen motion. The mesoscopic physics at play in 2DEGs is very rich and includes the integer and fractional quantum Hall effects,
quantum localization and double-barrier tunneling. One of the main building blocks of mesoscopic devices
is the so-called ``quantum point contact'' (QPC)~: a constriction of the 2D electronic fluid created by the
electrostatic potential of a gate. Tuning the potential allows to control the shape of the electron droplet.
Devices like interferometers involve typically several QPCs. When used in the quantum Hall regime they allow
manipulation of the edge modes that propagate at the boundary of the sample. It is possible to increase the tunneling rate
between edges by inducing a pinching of the 2DEG. By measuring the noise characteristics of this geometry it has been
shown~\cite{Glattli97,Moty97} that quasiparticles at filling factor 1/3 in the fractional quantum Hall (FQH)
regime have fractional charge 1/3 as expected
from Laughlin's theory~\cite{Laughlin83}. 

The simplest description of these phenomena is by an effective theory
that keeps only the low-energy edge modes that are generically present in FQH liquids~\cite{AHM90,Wen90,Wen92,Wen95}.
In this framework the simplest FQH fluids at filling factor $\nu=1/m$ have low-energy excitations described by a free chiral boson
theory at each edge. Complicated samples with multiply-connected geometry involve in general several such modes even
if the electron fluid stays at $\nu=1/m$. When the filling factor is not simple, for example in the Jain sequence
$\nu =p/(2p+1)$, edge states are described by several interacting chiral bosons. 
Detailed understanding of these modes is still 
controversial~\cite{Jolad07,Jolad2010a,Jolad2010b,Sreejith2011,Chang03,Goldman01,Wan05,Palacios96,Papa04,Zulicke03}.
Even more intriguing is the case
of the fraction $\nu =5/2$ which is believed to be described by the Moore-Read~\cite{Moore91} Pfaffian state (or its particle-hole
conjugate dubbed the antipfaffian). This elusive FQH state has a fermionic Majorana edge mode in addition to a chiral bosonic mode
and interferometers have been proposed to manipulate the associated quasiparticles and reveal
their non-Abelian statistics~\cite{RMP}. Typical devices like Fabry-Perot interferometers involve several QPCs.
As a function of the gate voltage there are typically two limiting regimes. When the voltage is small the pinching
of the 2DEG leads to a smaller distance distance between edges and hence tunneling of quasiparticles is possible
through traversal of the bulk liquid. When the voltage is very large one may reach the so-called pinch-off regime where the droplet
is now cut into two separate islands and only electron tunneling between them is allowed. The theoretical treatment
of these two situations can be done by adding ad-hoc operators to the effective bosonic field theory, encapsulating all microscopic
tunneling details into a few coupling constants. The quasiparticle creation operator has then to be expressed in terms
of the effective Bose degrees of freedom, as is the case for the electron operator.
Many experimental features can be explained in this framework~\cite{Chang03}. Nevertheless it is important to have a microscopic understanding
of the phenomena. Previous numerical studies of microscopic models have used the disk 
geometry~\cite{Wan06,Chen08,Wan08,Chen09,Hu09,Hu11} to evaluate tunneling matrix elements 
for the fillings $\nu=1/3$ as well as $\nu=5/2$. It is also possible up to some extent to identify the edge mode
quantum numbers and measure velocities.

In this paper we study the FQH effect in a QPC by using the cylinder geometry~\cite{Chamon94,Soule2012,DMRG}. 
In the Landau gauge one can impose periodic boundary conditions along one direction and keep the other one free.
It is then possible to perform exact diagonalization (ED) studies of various Hamiltonians for small numbers of particles.
This gauge also allows the addition of a parabolic confining potential to lift the edge mode degeneracy.
We first use a hard-core interaction two-body interaction to create a $\nu=1/3$ FQH droplet. The spectrum of low-lying modes
can be analyzed in terms of a Luttinger model~\cite{Haldane81}. While this is known for the case of a positive potential~\cite{Soule2012}
we show that if the harmonic potential is reversed (with hard walls to keep it bounded from below) then the fluid
separates in two droplets and that the two disconnected edges again combine in a non-chiral Luttinger liquid.
The study of the spectrum allows us to identify the zero modes of the effective bosonic theory and their role
in the description of electron tunneling. Measurement of the Luttinger exponent reveals the duality between
quasiparticle and electron properties as expected from theory~\cite{KF92,KF92L,KF94,KF95,KFP94}.

We also study the Moore-Read Pfaffian state in the QPC geometry by using the special three-body hard-core
Hamiltonian whose ground state is the Pfaffian. The mode counting reflects the fractionalization of the charge in units
of $e/4$. In the spectrum we are able to identify the various bosonic and fermionic excitations that form conformal towers
as expected from conformal field theory (CFT). While there is a set of low-energy states that corresponds to
almost rigid translation of the incompressible droplet this is not the only possibility and we also find a series of states
that are generated by transfer of a $e/4$ quasiparticle through the droplet. The associated conformal towers
now are described in terms of the spin scaling field in the Ising CFT. 
It is an important feature of the cylinder geometry that the non-Abelian sector appears in a natural way as a function
of the total momentum of the ground state and does not require tweaking extra potentials to locate a quasiparticle
at the origin of a disk or a the pole of a sphere.
The low-lying energies of the CFT towers
can be analyzed to find the charge Luttinger parameter as in the plain Laughlin liquid case but lead also to the measurement of
the non-trivial scaling dimensions of the Ising fields. 
While ED studies are limited to numbers of particles too small to extract the scaling dimensions, we introduce
a Monte-Carlo method based on the exact formulas for Pfaffian edge states in all sectors to compute
reliably the scaling dimension of spin field $\sigma$ as well as of the Majorana field $\theta$.
The bosonic Pfaffian at $\nu=1$ share the same spectral features.
For the special hard-core Hamiltonians that generate exactly the wavefunctions we find unexpectedly that the Bose and Fermi
velocities are equal or almost equal in the thermodynamic limit which implies that the special interactions have an extended
symmetry.

In section \ref{geo}, we present the cylinder geometry and discuss the introduction of a parabolic potential.
In section \ref{Lut}, we add a parabolic potential well to lift the degeneracy of the edge states and compute
the Luttinger parameter $g$ in the simple case of a principal Laughlin $\nu =1/m$ in the two possible geometries
of the QPC, the weak constriction regime and the pinch-off regime. In section \ref{pfaffiandisk} we explain the basics
of the Moore-Read Pfaffian edge modes in the disk geometry. Then we explore the conformal towers in the cylinder
geometry in section \ref{cylindersection}. Finally we present a Monte-Carlo method to obtain accurate measurements of scaling dimensions
in section \ref{MC}.
Our conclusions are presented in section \ref{conclude}.

%****************************************************************************************
\section{The QPC in the cylinder geometry}
\label{geo}

To study FQH physics on the cylinder one has first to use the Landau gauge which
is compatible with periodic boundary conditions in one space direction. 
We take the vector potential as $A_x=0$ and $A_y=B x$ and
the one-body eigenstates in the lowest Landau level (LLL) are given by~:
\begin{equation}
\phi_{n} (x,y)= 
\frac{1}{\sqrt{L  \sqrt{\pi}}}
{\mathrm{e}}^{-\frac{1}{2}{(x-k)^2}}
{\mathrm{e}}^{iky}=
\frac{1}{\sqrt{L  \sqrt{\pi}}}
{\mathrm{e}}^{-\frac{2\pi^2}{L^2}{ n^2}}
Z^n \,\,
{\mathrm e}^{\displaystyle{- {x^2}/{2 }}},
\quad
 Z\equiv {\mathrm e}^{\frac{2\pi}{L}(x + i y) }\, .
\end{equation}
Here we have set the magnetic length $\ell =\sqrt{\hbar c/eB}$ to unity.
The periodic direction $y$ has a length $L$ and the momentum $k$ is quantized $k=2\pi n/L$
with $n$ an integer (which can be negative or positive). All the states $\phi_n$ are degenerate and constitute the basis
on which we construct the Fock space.  If we add a potential energy $V(x)$ invariant by translation
along the cylinder periodic direction $y$ then it appears in the Hamiltonian through its matrix element~:
\be
{\tilde V}_n = \frac{1}{L\sqrt{\pi}}\int dx \,\,{\mathrm e}^{\displaystyle{- {(x-2\pi n/L)^2}}} V(x).
\label{potential}
\ee
To deal with a finite-dimensional Fock space we impose a cut-off on the values
of the integer $n$. This procedure allows ED of the Hamiltonian by linear algebra
techniques standard in the field of strongly correlated systems. This type of cut-off physically resembles a cut-off in real space 
only in the limit when the length $L\rightarrow 0$. Here the orbitals are widely separated by more than the magnetic length
and any reasonable confining potential gives rise to such a boundary condition. In the opposite limit
$L\rightarrow\infty$ there is \textit{no} non-oscillating potential leading to  a hard cut-off in reciprocal space.
This leads ultimately to the appearance of spurious low-energy modes that do not belong to the true FQHE problem~\cite{Soule2012}.
However if the keep an aspect ratio reasonably close to unity, then the physics of the FQHE is present in this geometry.
The limit $L\rightarrow 0$ is known as the Tao-Thouless limit (TT)~\cite{Tao83,Thouless84,Chui86,BHHK,BK,Seidel05}. 
The problem becomes closer to an electrostatic problem
and many features of the FQHE can be analyzed simply. However in the TT limit the ground state becomes a Slater determinant
with a trivial entanglement spectrum. Nevertheless the fusion rules for quasiparticles still remain complete~\cite{Ardonne}. We show
in this paper that this is also an appropriate limit to identify the so-called conformal towers of excited states.

Generically the two-body interaction projected in the LLL can be written as a function of projection operators
of the state of relative angular momentum $m$ for each pair of particle~:
\be
{\mathcal H}_{int}=\sum_{i<j} \sum_m V_m {\hat P}_{ij}^{(m)} ,
\label{pseudos}
\ee
where the coefficients $V_m$ are the so-called~\cite{Haldane81} Haldane pseudopotentials. While originally defined in the disk
geometry they can be extended to the cylinder case as shown by Rezayi and Haldane~\cite{RH94}. Spinless fermions are only sensitive
to odd values of $m$ and of particular interest is the extreme hard-core model for which the only nonzero
pseudopotential is $V_1$. In this peculiar case the Laughlin wavefunction is the exact unique zero-energy ground state.
A fictitious potential reproducing an arbitrary set of $V_m$s is given by ${\tilde V}(q)=\sum V_m L_m(q^2)$
where the $L_m$s are the Laguerre orthogonal polynomials. It can be then inserted in the second-quantized expression
of the Hamiltonian~:
\be
{\mathcal H}_{int}=\frac{1}{2}\sum_{\{m_i\}}\,
{\mathcal A}_{m_1,m_2,m_3,m_4}\,
 c^\dag_{m_1} c^\dag_{m_2} c_{m_3} c_{m_4} ,
\ee
with the matrix elements given by~:
\be
{\mathcal A}_{m_1,m_2,m_3,m_4}=
\frac{1}{2L}\int \frac{dq}{2\pi}\,\,
{\tilde V}(q,\frac{2\pi}{L}(m_1-m_4))\,
{\mathrm e}^{\displaystyle{-q^2/2+2i\pi q (m_1-m_3)/L}}\,
{\mathrm e}^{\displaystyle{- {2\pi^2 (m_1-m_4)^2}/{L^2 }}},
\ee
where ${\tilde V}(q_x,q_y)$ is the ordinary Fourier transform of the potential.
In the present work we use a potential that produces only $V_1$ and $V_3$
given by~:
\be
{\tilde V}(q_x,q_y)\equiv
{\tilde V}({\bf q})=-(V_1+3V_3){\bf q}^2+\frac{3}{2}V_3 {\bf q}^4 -\frac{1}{6}V_3 {\bf q}^6,
\ee
where we set ${\bf q}^2=q_x^2+q_y^2$. The coefficients for the third pseudopotential can be checked
by verification that the Hamiltonian has a unique zero-energy ground state for filling factor $\nu =1/5$
obtained by imposing the (Laughlin) relation between the number of orbitals and the number of electrons.

If we want to study the Pfaffian state then one needs to consider a special three-body Hamiltonian
with derivative interactions whose exact zero-energy ground-state is given by the Pfaffian wavefunction
introduced by Moore and Read~\cite{Moore91}~:
\be
\mathcal{H}_3=-\sum_{i<j<k}\mathcal{S}_{ijk}\Delta_i^2\Delta_j 
\delta^2({\bf r}_i-{\bf r}_j)\delta^2({\bf r}_j-{\bf r}_k)
\label{H3},
\ee
where $\mathcal{S}$ stands for symmetrization.
To perform ED studies the Fock space should be finite-dimensional. This is enforced by
restricting the allowed orbitals to a finite range of momenta~: the integer $n$ indexing the one-body wave functions
takes only $2K_{max}+1$ values. This allowed range can be translated at will as long as there is complete
translation invariance along the cylinder.
For example if we take $0\leq n\leq 2K_{max}$ then the Z powers are positive and formal expressions of
first-quantized wavefunctions are closest to the disk geometry. Centering the range of $n$ values at zero momentum
is more natural if we add a parabolic potential to mimic a realistic QPC geometry.
To create a droplet of Laughlin-type fluid at filling 1/3 one chooses $2K_{max}=3(N-1)$ while the Pfaffian
requires the adjustment $2K_{max}=2N-3$ exactly as in the spherical geometry. Adding orbitals beyond this
reference number leads to the appearance of more zero-energy states that are the gapless edge states of the
FQHE fluid. In the cylinder geometry there are \textit{two} counterpropagating edges that combine into
non-chiral effective theories. To reveal their precise content it is convenient both physically and practically
to add an extra parabolic potential well along the axis of the cylinder which is invariant under $y$ translations so that
the total momentum remains a good quantum number~: $V=\beta \sum_k k^2 n_k$.
We define the conserved total momentum as $K_{tot}$
whose values are integers or half-integers that can be used to label the many-body eigenstates
(this momentum is thus in units of $2\pi/L$).

%%%%%%%%%%%%%%%%%%%%%%%%%%%%%%%%%%%%%%%%%%%%%%%%%%%%%%%%
\begin{figure}[htb]
\includegraphics[width=0.4\columnwidth]{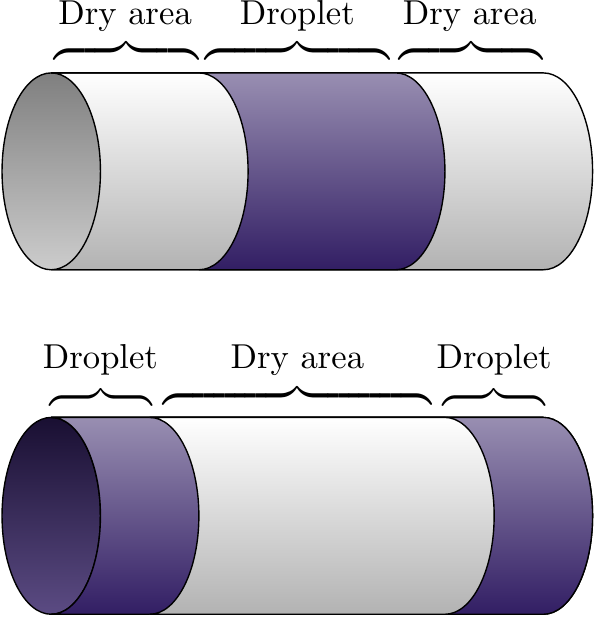}
\caption{The location of the quantum Hall droplet according to the value of the potential along the axis of the cylinder.
Top panel~: the potential is convex leading to a single droplet (blue area) centered in the middle of the cylinder.
Bottom panel~: with a concave potential it is more favorable to split the liquid into two occupied areas close
to the hard walls. The edge modes then reside at the boundary between the droplet(s) and the dry area(s).}
\label{NiceCylinders}
\end{figure}
%%%%%%%%%%%%%%%%%%%%%%%%%%%%%%%%%%%%%%%%%%%%%%%%%%%%%%%%

%%%%%%%%%%%%%%%%%%%%%%%%%%%%%%%%%%%%%%%%%%%%%%%%%%%%%%%%
\begin{figure}[htb]
\includegraphics[width=0.4\columnwidth]{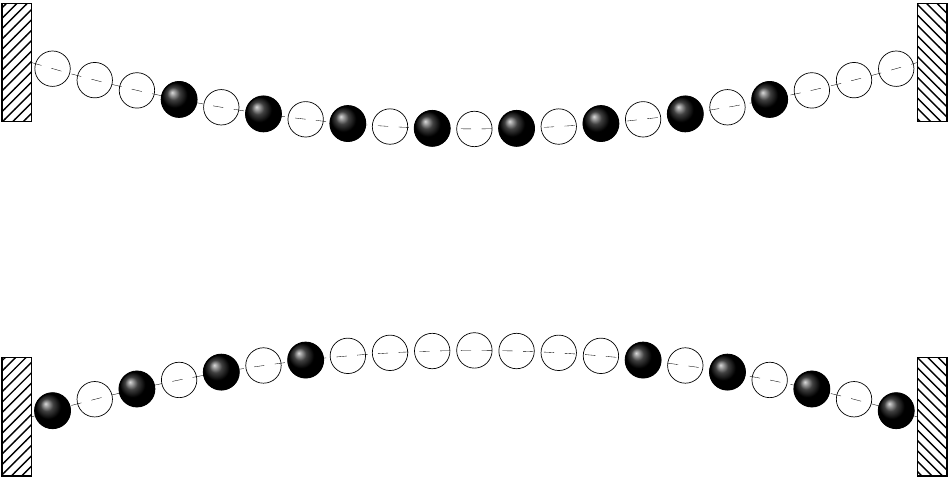}
\caption{The root configuration of a bosonic Laughlin state at $\nu =1/2$ in the two possible configurations mimicking
the QPC geometry. Top view shows the middle droplet while the bottom view shows the configuration with two spatially separated droplets.}
\label{Root}
\end{figure}
%%%%%%%%%%%%%%%%%%%%%%%%%%%%%%%%%%%%%%%%%%%%%%%%%%%%%%%%

\begin{figure}[htb]
\includegraphics[width=0.8\columnwidth]{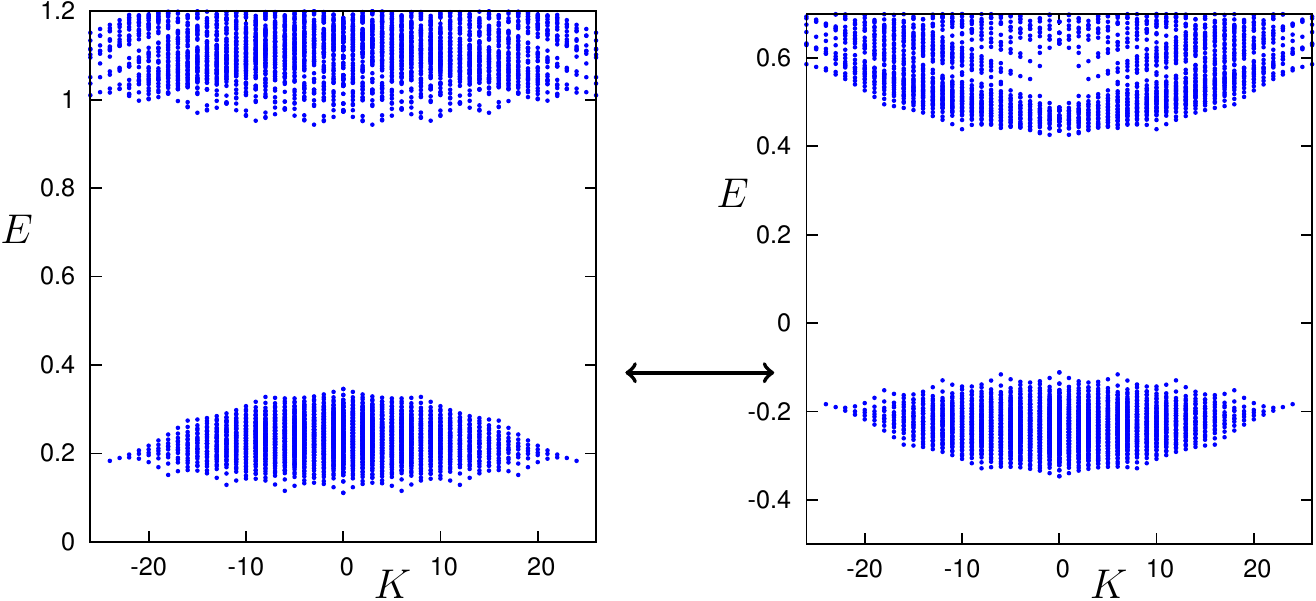}
\caption{The two QPC configurations in the limit of a small parabolic potential. 
Here we plot the spectrum of $N=6$ electrons in orbitals describing the $\nu=1/3$ Laughlin state
and edge excitations.Left panel: single droplet regime
where the cusps in the low-lying states are related to quasiparticle transfers. Right panel: we change the sign of the confining
potential to switch to the pinch-off regime. Now the cusps are due to electron tunneling. Since the potential is taken to be weak,
first-order perturbation theory means that one just reverse the manifold of edge states.}
\label{QPCconfig}
\end{figure}

If the coefficient $\beta$ is taken to be positive then a quantum Hall droplet will stay centered in the middle of the cylinder
and, provided we allow enough extra orbitals, there will be two edges that can combine. This is is displayed in Fig.(\ref{NiceCylinders})
and Fig.(\ref{Root}).
If we now use a negative $\beta$ value leading to a inverted potential then the droplet splits into two separate
chunks each of them using the more energetically favorable orbitals closest to the hard wall in $k$-space, provided
that the potential well is deep enough. In the following studies we use a potential small enough so that there is always
a large separation between bulk and edge states~: see Fig.(\ref{QPCconfig}) for a typical situation.

Since we use a parabolic external potential $V=\beta \sum_k (k-k_0)^2 n_k$, it  satisfies~:
\be
[V,U]= \beta U^p
\sum_k \{(k-k_0+p)^2-(k-k_0)^2\} n_{k} = \beta U^p (2 p K + p ^2 N ) ,
\label{tr1}
\ee
where $U\propto \prod_i Z^i$ is the translation
operator of one orbital along the x axis, and $K=\sum_k (k-k_0) n_k$ the momentum operator with the convention that the
ground state wavefunction has zero momentum. Therefore, if $| \psi \rangle $ is an eigenstate of our total
Hamiltonian $\mathcal H_{int} + V$ with energy $E$, $N$ particles and momentum $K$, then $U^p | \psi \rangle$ is also an
eigenstate with energy and momentum : 
\be
E'=\beta (2 p K + p^2 N), \quad K'=K+pN\, .
\label{tr2}
\ee
So, provided there are no hard walls or if the walls are far enough (large $K_{max}$), it is possible to deduce the whole spectrum from the
sector with $K=0,\ldots,N-1$. It can be used in an approximate way if we add enough orbitals beyond that required by the flux-number of electrons that
defines our FQH state.

%%%%%%%%%%%%%%%%%%%%%%%%%%%%%%%%%%%%%%%%%%%%%%%%%%%%%%%%
%%%%%%%%%%%%%%%%%%%%%%%%%%%%%%%%%%%%%%%%%%%%%%%%%%%%%%%%

\section{Luttinger liquid and the Abelian QPC}
\label{Lut}

%%%%%%%%%%%%%%%%%%%%%%%%%%%%%%%%%%%%%%%%%%%%%%%%%%%%%%%%%%

\begin{figure}[htb]
\begin{tikzpicture}
\draw [>=latex,->] (0,0) -- (10,0);
\draw [>=latex,->] (0,0) -- (0,5);
\draw (10,0) node[below right] {$K_{tot}$};
\draw (0,5) node[left] {E};
\draw (0,0) node[below] {$0$};
\draw (4,0) node[below] {$2k_F$};
\draw (8,0) node[below] {$4k_F$};
%
% K=0 excitations
\draw (0,0) node{$\bullet$};
\draw (-0.5,1.0) node{$\bullet$};
%\draw (0,1.0) node{$\bullet$};
\draw (+0.5,1.0) node{$\bullet$};
\draw (-1.0,2.0) node{$\bullet$};
%\draw (-0.5,2.0) node{$\bullet$};
\draw (+0.0,2.0) node{$\bullet$};
%\draw (+0.5,2.0) node{$\bullet$};
\draw (+1.0,2.0) node{$\bullet$};
\draw (-1.5,3.0) node{$\bullet$};
%\draw (-1.0,3.0) node{$\bullet$};
\draw (-0.5,3.0) node{$\bullet$};
%\draw (+0.0,3.0) node{$\bullet$};
\draw (+0.5,3.0) node{$\bullet$};
%\draw (+1.0,3.0) node{$\bullet$};
\draw (+1.5,3.0) node{$\bullet$};
%
% 2kF excitations
\draw (4-0,0+0.2) node{$\bullet$};
\draw (4-0.5,1.0+0.2) node{$\bullet$};
%\draw (4+0,1.0+0.2) node{$\bullet$};
\draw (4+0.5,1.0+0.2) node{$\bullet$};
\draw (4-1.0,2.0+0.2) node{$\bullet$};
%\draw (4-0.5,2.0+0.2) node{$\bullet$};
\draw (4+0.0,2.0+0.2) node{$\bullet$};
%\draw (4+0.5,2.0+0.2) node{$\bullet$};
\draw (4+1.0,2.0+0.2) node{$\bullet$};
\draw (4-1.5,3.0+0.2) node{$\bullet$};
%\draw (4-1.0,3.0+0.2) node{$\bullet$};
\draw (4-0.5,3.0+0.2) node{$\bullet$};
%\draw (4+0.0,3.0+0.2) node{$\bullet$};
\draw (4+0.5,3.0+0.2) node{$\bullet$};
%\draw (4+1.0,3.0+0.2) node{$\bullet$};
\draw (4+1.5,3.0+0.2) node{$\bullet$};
%
% 4kF excitations
% 
\draw (8-0,0+0.8) node{$\bullet$};
\draw (8-0.5,1.0+0.8) node{$\bullet$};
%\draw (8+0,1.0+0.8) node{$\bullet$};
\draw (8+0.5,1.0+0.8) node{$\bullet$};
\draw (8-1.0,2.0+0.8) node{$\bullet$};
%\draw (8-0.5,2.0+0.8) node{$\bullet$};
\draw (8+0.0,2.0+0.8) node{$\bullet$};
%\draw (8+0.5,2.0+0.8) node{$\bullet$};
\draw (8+1.0,2.0+0.8) node{$\bullet$};
\draw (8-1.5,3.0+0.8) node{$\bullet$};
%\draw (8-1.0,3.0+0.8) node{$\bullet$};
\draw (8-0.5,3.0+0.8) node{$\bullet$};
%\draw (8+0.0,3.0+0.8) node{$\bullet$};
\draw (8+0.5,3.0+0.8) node{$\bullet$};
%\draw (8+1.0,3.0+0.8) node{$\bullet$};
\draw (8+1.5,3.0+0.8) node{$\bullet$};
\draw [dashed,domain=0:10] plot (\x,{0.0125*\x*\x}); 
\end{tikzpicture}

\caption{The spectrum of a free non-chiral boson theory with finite radius. There are sectors at
$K=0,2k_F,4k_F,\dots$ that are similar to the spectrum of a non-compact boson. The extra excitations associated
with non-trivial windings leads to copies of the $k=0$ spectrum shifted quadratically in energy. The finite-size
spectrum of free fermions is the same. The Luttinger parameter value is encoded in the parabolic shift of the energy. The special
spectra above each extremal point at $2nk_F$ are the so-called conformal towers of the bosonic CFT.}
\label{BosonSpectrum}
\end{figure}
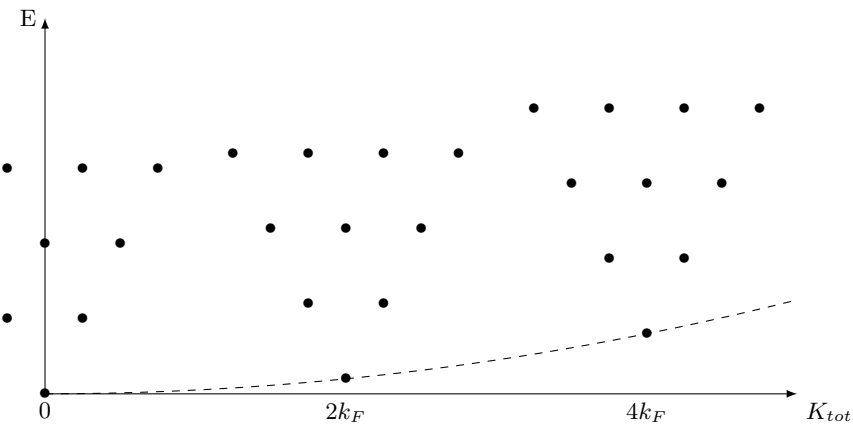

The Lagrangian of the chiral Luttinger model describing edge excitations of a Hall sample with two counterpropagating edges
around an incompressible FQH liquid at filling $\nu =1/m$ is given by~\cite{Wen95,Chang03}:
\be
{\mathcal L}={\mathcal L}_R+{\mathcal L}_L=\frac{1}{4\pi g}\{
\partial_x\phi_R (i\partial_\tau+v\partial_x)\phi_R+
\partial_x\phi_L (-i\partial_\tau+v\partial_x)\phi_L\},
\label{LutLR}
\ee
where the Luttinger parameter $g$ is equal to the filling factor $g=\nu$.
This effective theory does not include edge interactions and is presumably valid when there is a large spatial separation
between them. The Hamiltonian that follows can be written as~:
\be
\mathcal{H}=\frac{v}{4\pi g}\int_0^L dx\, \{(\partial_x \phi_L)^2 + (\partial_x \phi_R)^2\}.
\label{HLR}
\ee
To expand this Hamiltonian into Fourier modes, one has to keep track of the zero modes that comes from the periodicity
of the bosonic fields~:
\be
\mathcal{H}=\frac{\pi v}{g L}(N_L^2+N_R^2)+ \sum_{q\neq 0}v|q| b^\dag_q b_q\, ,
\label{zeromodes}
\ee
where $b_q^\dag , b_q$ are the Fourier mode creation/annihilation operators and $N_{L,R}$ are the winding numbers of
the bosonic fields along the edge~:
\be
\phi_{L,R}(x+L)-\phi_{L,R}(x)=2\pi N_{L,R}.
\ee
Quasiparticle states corresponds to fractional values of these parameters $N=n\nu$ with $n$ integer while
electron states are generated with $N=n$. This can be seen by computing the electric charge associated
with the winding through $\rho =\partial_x \phi /2\pi$.
The momentum operator of this theory has a zero mode contribution in addition to the phononlike part~:
\be
K=k_F J +\sum_{q\neq 0}q b_q^\dag  b_q \, ,
\ee
where the even integer $J$ counts essentially the particle-hole excitations across the pseudo-Fermi surface
with Fermi momentum $k_F$.
Finally we note that the number of states at a given energy for a given chirality is determined by the partitions
of unity and this
can be obtained easily by expanding the character~:
\be
\chi_{B} (q) =  \prod_{n=1}^{\infty} \frac{1}{\left( 1 - q^{n} \right)}.
\label{Char}
\ee
This character is that of a free boson CFT.
The level scheme we expect from the Luttinger picture is shown in Fig.(\ref{BosonSpectrum}).

\begin{figure}[htb]
\includegraphics[width=0.6\columnwidth]{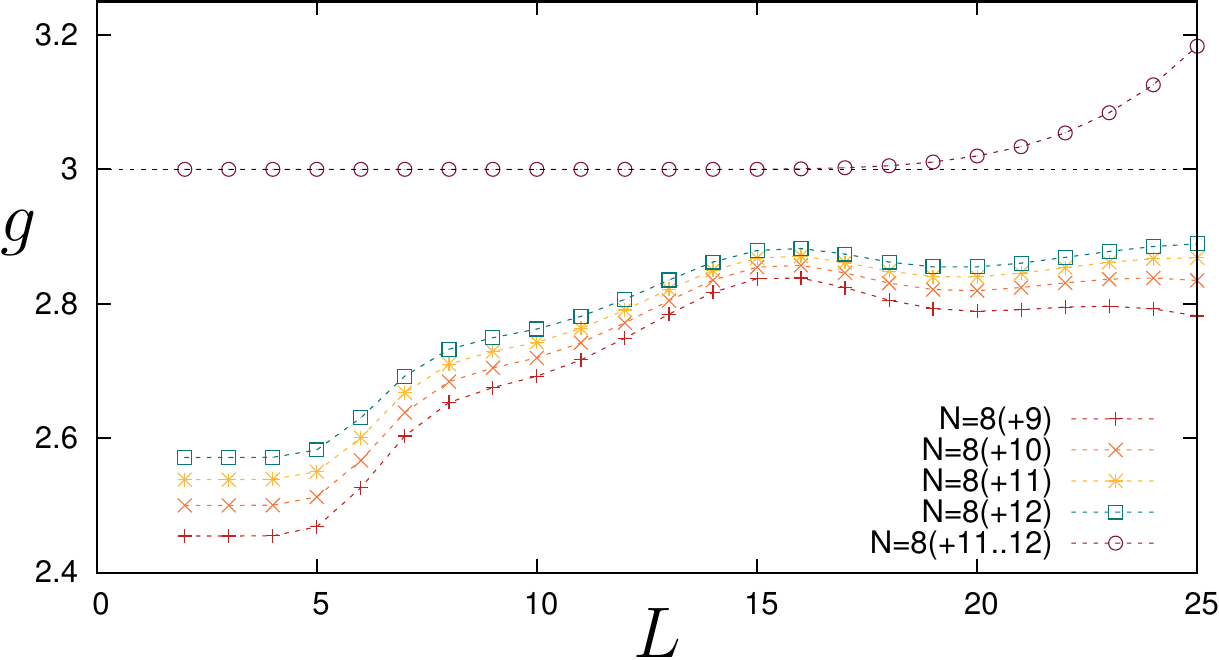}
\caption{Numerical estimation of the Luttinger parameter in the electron-tunneling configuration for two $\nu=1/3$ 
spatially separated FQH
droplet. The first four curves at the bottom are obtained from the edge spectrum for $N=8$ with fixed number of extra
orbitals. The topmost curve, circle markers, is an estimation using the finite size scaling explained in section
\ref{MC} with $11$ and $12$ extra orbitals. We also find that adding a small $V_3$ pseudopotential does not
change the value of the exponent.}
\label{Dualexp}
\end{figure}

If we now compute by exact diagonalization the spectrum of an electron system with a small quadratic potential
then we find a characteristic Luttinger-like spectrum with arches interpolating between quasi-ground states
separated by momenta $K=\pm N_e, \pm2N_e,\pm3N_3,\dots$. By adiabatically following these quasi-ground states
when sending the potential to zero, we can identify their root configuration. In fact 
they are obtained by a global uniform translation of the Laughlin state from Eqs.(\ref{tr1},\ref{tr2}). For neighboring quasi-ground states the shift
is by exactly one orbital. This operation can be realized by threading the cylinder with a thin solenoid and increasing
adiabatically the flux by one quantum. This is the celebrated Laughlin argument for quasiparticle creation.
Electron are then pushed along the axis of the cylinder by exactly one orbital at the end the process.
This is equivalent to the transfer of a fractionally charged quasiparticle from one edge of the system to the other edge.
These quasiparticle transfers generate the whole set of extremal states that are prominent in Fig.(\ref{QPCconfig}). On top of these global
excitations we can of course excite phononlike density modes with two possible chiralities corresponding to the two
edges of the quantum Hall droplet.
These excitations are created by acting with the $b^\dag_q$ Fourier mode operators.  
We identify the modes that are above each of the extremal states as phonon modes. The counting we find
in ED studies is always compatible with the bosonic counting rule Eq.(\ref{Char}).
From Eq.(\ref{zeromodes}) we see that extracting the energies of the parabola of extremal states
lead to an estimate of the Luttinger parameter, provided we have an independent measurement of the velocity $v$ of phonon excitations.
This strategy was already applied in the weak constriction regime~\cite{Soule2012}. Here we note that it may also be applied
in the complementary case of the QPC in the pinch-off regime where the fluid is now separated into two droplets.
Indeed the global contribution to the Hamiltonian Eq.(\ref{HLR}) can be written as~:
\be
\mathcal{H}_{global}=\frac{\pi v}{2\nu L}(J^2+N^2),
\ee
where $J=N_L-N_R$ ($N=N_L+N_R$) is an integer that changes by two units for each transfer of an electron from the left droplet to the right droplet.
This integer $J$ indexes the extremal low-lying states of the Luttinger spectrum in Fig.(\ref{QPCconfig}) in the split QPC regime.
Fitting the global parabolic envelope of the spectrum in this regime leads then to the \textit{dual} exponent $g=1/\nu =3$ as expected from the
standard duality~\cite{Wen95} between electrons and quasiparticles. Even with small system sizes we obtain a precise estimate of $g$~: 
see Fig.(\ref{Dualexp}).

%%%%%%%%%%%%%%%%%%%%%%%%%%%%%%%%%%%%%%%%%%%%%%%%%%%%%%%%
%%%%%%%%%%%%%%%%%%%%%%%%%%%%%%%%%%%%%%%%%%%%%%%%%%%%%%%%

We finally note that the tower structure of the principal Laughlin fluid can also be revealed by computing the entanglement 
spectrum~\cite{Sterdyniak,Sterdyniak2,Lauchli2010}.
However this approach does not lead to estimates of the boson radius.

%%%%%%%%%%%%%%%%%%%%%%%%%%%%%%%%%%%%%%%%%%%%%%%%%%%%%%%%%%%
\section{the pfaffian edge states in the disk geometry}
\label{pfaffiandisk}

We now turn to the study of the model wavefunction known as the Moore-Read Pfaffian~\cite{Moore91}.
We first recall for completeness the knowledge about edge modes in the disk geometry before turning to the analysis
in the cylinder geometry with confining potential modeling a constriction in the $\nu =5/2$ state.
The formula for the Pfaffian state in the disk geometry is~:
\be
\Psi_{MR}(z_i)=\mathrm{Pf}(\frac{1}{z_i-z_j})\prod_{i<j}(z_i-z_j)^m \, ,
\label{pfaffianMR}
\ee
where the symbol Pf stands for the Pfaffian of a matrix~:
\be
\mathrm{Pf}(A_{ij}) =\sum_\sigma \epsilon_\sigma
A_{\sigma(1)\sigma(2)}  \dots  A_{\sigma(N-1)\sigma(N)}\, ,
\ee
where $\epsilon_\sigma$ is the signature of the permutation $\sigma$.
The corresponding filling factors are $\nu=1$ in the Bose case and $\nu=1/2$ in the Fermi case. In some 2DEGs with very high mobility
there is an incompressible state in the second LL with total filling $5/2=2+1/2$ which is a strong candidate to be described by this 
wavefunction~\cite{RMP}.
Bosons with low-energy scattering in the s-wave are also an interesting candidate for the $m=1$ state above~\cite{Regnault03,Chang05}.
This model wavefunction can be generated as the zero-energy eigenstate of the three-body hard-core Hamiltonian Eq.(\ref{H3}). 
The edge modes of these fractions have been studied by Wen~\cite{wen93} and Milovanovic and Read~\cite{MR96}.
They include the charged modes that we have already discussed in the case of the elementary Laughlin fractions at filling
$\nu =1/m$. They are generated by multiplying the wavefunctions by a symmetric polynomial. The counting is given by 
a free bosonic mode Eq.(\ref{Char}).  However this is not the whole story~: there are also extra neutral modes 
that can be described by a Majorana fermion theory. 
From the point of view of the wavefunction, they are generated by removing some of the $1/z$ pairing-type factors and replacing them by powers
of the coordinates~:
\be
\Psi_{n_i}=\mathcal{A}(z_1^{n_1} z_2^{n_2}\dots z_F^{n_F}\frac{1}{z_{F+1}-z_{F+2}}
\dots\frac{1}{z_{N-1}-z_N})\prod_{i<j}(z_i-z_j)^m \, ,
\ee
where $\mathcal{A}$ means antisymmetrization. The non-negative and distinct  integers $0\leq n_1<\dots< n_F$
can be interpreted as the occupation numbers of a  fermion. Note that in this formula the number of electrons $N$ can be odd or even 
(contrary to our basic Pfaffian definition Eq.(\ref{pfaffianMR})) but the difference
$N-F$ should be even. This set of states can be counted by a
 Majorana-Weyl field theory provided we use antiperiodic boundary conditions on the field operator $\theta$.
This field theory can be written in a Lagrangian language~:
\be
\mathcal{L}_M = i\theta (\partial_t - v_n \partial_{x}) \theta\, ,
\ee
where $v_n$ is the velocity of the fermionic modes. It is nonzero only in the presence of an external potential such
as the one created in a QPC. The antiperiodic sector $\theta(x+L)=-\theta(x)$ is called the Neveu-Schwarz (NS) sector.
These edge modes are essentially unaffected by the presence of Abelian (i.e. Laughlin) quasihole in the bulk liquid, here
in the interior region of the circular droplet. However the Pfaffian state of matter also include quasiholes with $e/4$ electric charges
and such quasiholes when present in an odd number in the bulk do change the physics of the edge modes.
This case is called the ``twisted'' sector~\cite{MR96,Fendley06,Fendley07}. Again these states with $F$ 
fermion edge modes require $N-F$ to be even.
They are properly counted by the fermion field theory above provided we impose now periodic boundary conditions
on the field operator $\theta(x+L)=+ \theta(x)$. This is the Ramond (R) sector. The boundary conditions imposed
on the fermion field operator give rise to different quantization conditions on momenta and hence to different counting
of edge modes. In the R (respectively NS) sector,
the Majorana fermionic modes are $\theta_n$ with $n$ integer (respectively $n$ half-integer). 

In the disk without $e/4$ quasiholes in the bulk, for even numbers of electrons we have even numbers of NS Majorana fermions
while for odd number of electrons we have odd numbers of NS fermions. This is the case studied by Wen~\cite{wen93}.
However, his ED
studies in the disk geometry missed the non-Abelian sector associated with the mere existence~\cite{Moore91} of fractional charges $e/4$.
If we add an extra non-Abelian quasihole for example at the center of disk by tuning an external potential~\cite{Hu11}
we now enter the R sector with again even/odd number of Majorana fermions for even/odd number of electrons.
In these studies one has also to consider the elementary charge mode excitations described by the single boson that is already
present in simple principal liquids at $\nu=1/m$. In general the velocities of the edge modes $v_n$ and $v_c$ are not equal.
 In fact studies in the disk geometry
have found estimates of these velocities that are quite different~\cite{Hu11}. They were performed for interactions that are a linear
combination of the hard-core three-body interaction Eq.(\ref{H3}) and Coulomb interaction. 
% chiral table of states
\begin{table}
\begin{tabular}{|l|l|l|}\hline
 $I$  & 1 0 1 1 2 2 3 3 5 & NS even\\ \hline
$\theta$         & 1 1 1 1 2 2 3 4 5 & NS odd\\ \hline
$\sigma,\mu$ & 1 1 1 2 2 3 4 5 6 & R even/odd\\ \hline\hline
$\phi +I $ & 1 1 3 5 10 16  28& NS even \\ \hline
$\phi +\theta$ & 1 2 4 7 13 21 35 & NS odd \\ \hline
$\phi + \sigma,\mu$ & 1 2 4 8 14 24 40 & R even/odd \\ \hline
\end{tabular} 
\label{chiraledge}
\caption{The edge state counting in the disk geometry. Top table is the number of states of the various sectors of the
Ising CFT. Bottom table include the bosonic excitations assuming that the velocities of the bosons and fermions are the same.}
\end{table}
The mode counting can be performed by elementary means by expanding Bose and Fermi fields into Fourier modes
and populating them according to the constraints. 

However it is important to realize that the Pfaffian universality class~\cite{Moore91}
is built upon a 2D CFT which is the product of an Ising model CFT times a free boson with specific radius.
This definition as a CFT implies some very specific properties of the spectrum of excited states. Most notably we expect
from the existence of the Virasoro algebra of conformal transformation the appearance of so-called conformal towers of states~\cite{ginsparg}.
In addition to the fermionic Majorana field,
the Ising CFT also contains the spin field $\sigma$
which cannot be expressed locally in terms of the  fermion field. 
As is known from the standard Kramers-Wannier duality of the Ising model, there is also a disorder spin field $\mu$ associated to 
$\sigma$ by duality. These fields,
contrary to the Majorana field, cannot be separated into chirality components. They are boundary-changing operators~\cite{ginsparg}
which act upon the boundary conditions of the  $\theta$ field operator~: NS$\longleftrightarrow$R. In the context of the FQHE the 
$\sigma,\mu$ operators are associated
with the presence of bulk $e/4$ quasiparticles that change the boundary conditions on the edge modes.
Finally, as in any CFT there is also the identity operator $I$.

\section{The pfaffian conformal towers on the cylinder}
\label{cylindersection}
It is known from the representation theory of general CFTs that
the spectrum of excited states are arranged into so-called conformal towers of states. In the 
cylinder geometry with left and right modes, to each primary operator $\Phi_{h,\bar{h}}$ with conformal dimensions $(h,\bar{h})$, 
corresponds an infinite set of states with (dimensionful) energies and momenta~:
\bea
 E &=& E_0 + \frac{2\pi v}{L}(h+\bar{h}+n+\bar{n}),\cr
P &=& P_0 + \frac{2\pi}{L}(h- \bar{h} + n-\bar{n}),
\label{generalCTower}
\eea
where $n,\bar{n}$ are integers, $v$ is the underlying velocity, and $E_0, P_0$ are 
respectively the ground-state energy and momentum.
There is simply a redefinition by a scale factor $2\pi/L$ with respect to the dimensionless
momentum $K_{tot}$ which we use in the remainder of the paper. 
In the Ising CFT, there are  four different sectors created by the action of primary operators on the vacuum.
On the cylinder geometry, we need to introduce two Majorana fermions for each edge:
$\theta_L$ and  $\theta_R$. The energy operator $\epsilon$ with conformal dimension 
$h=\bar{h}=1/2$ of the Ising CFT can be written directly in terms of these fermion fields~:
$\epsilon = i \theta_L \theta_R$. 
The four different sectors of the Ising CFT on the cylinder can be labeled by their fermionic content. 
We have the identity with $h=\bar{h}=0$ and energy tower with $h=\bar{h}=1/2$, which correspond 
to NS/even fermion states.
Then there is the Majorana sector with $h+\bar{h}=1/2$ given by NS/odd fermion numbers.
These two sectors are created by local Majorana operators $\theta_{L,R}$ .
Finally, we have the  the $\sigma$ and $\mu$ sectors with $h=\bar{h}=1/16$
corresponding to the R sector with respectively
 even and odd fermionic parity.  
The information on the counting of states  in each sector of the Ising CFT can be extracted from the 
so-called chiral Virasoro character~\cite{ginsparg}~:
\be
\chi_h (q) = q^{h - c/24} \sum_{n=0}^{\infty} d_h(n) q^{n},
\ee
where $c$ is the central charge, i.e. $c=1/2$ for the Ising CFT. 
In the following, we omit the overall $q^{-c/24}$ factor which is needed for modular invariance but not for the counting below. 
The counting of states in the disk geometry at momentum $\Delta P = 2\pi n/L$
is then given by the integers $d_h(n)$. For the Ising CFT, the different characters are given by \cite{ginsparg}:
\bea
\chi_{0} (q) &=& \frac{1}{2} \left[ \prod_{n=0}^{\infty} \left( 1 + q^{n+1/2} \right) + 
 \prod_{n=0}^{\infty} \left( 1 - q^{n+1/2} \right) \right] 
\nonumber \\
  \chi_{1/2} (q) &=& \frac{1}{2} \left[ \prod_{n=0}^{\infty} \left( 1 + q^{n+1/2} \right) - 
 \prod_{n=0}^{\infty} \left( 1 - q^{n+1/2} \right) \right] 
\nonumber \\
 \chi_{1/16} (q) &=& q^\frac{1}{16} \prod_{n=1}^{\infty} \left( 1 + q^{n} \right)  .
\label{chars}
\eea
By expanding the products, we get the counting of states of Table I in each conformal tower of the disk geometry
with one edge. By taking into account the Bose field contribution (see Eq. (\ref{Char})), we deduce the counting of the
edge states of the Pfaffian in the disk geometry (see lower table in Table I). 
In the cylinder geometry, the total number of states in the identity and energy sector is obtained through expansion of
$\chi_B^2 (\chi_0^2+\chi_{1/2}^2)$ while the Majorana fermion sector involves
$\chi_B^2 (2\chi_0\chi_{1/2})$ and the twisted $\sigma$ sector is related to $\chi_B^2\chi_{1/16}^2$.
By expanding the products, we get the counting of states, presented in Table II, in each conformal tower of the Pfaffian
at energy $\Delta E = 2\pi v n/L$.

% chiral table of states
\begin{table}
\begin{tabular}{|l|l|l|}\hline
 $\phi +I +\epsilon $ & 1 3 11 28 69 152 & NS even \\ \hline
$\phi +\theta_{L,R}$ & 2 6 18 44 104 222& NS odd \\ \hline
$\phi + \sigma,\mu$ & 1 4 12 32 76 168 & R even/odd \\ \hline
\end{tabular} 
\label{cylinderedge}
\caption{The edge state counting in the non-chiral cylinder geometry. This refers to energy level degeneracy
taking into account all allowed momenta. We have assumed that Fermi and Bose velocities are equal.}
\end{table}

\begin{figure}[htb]
\includegraphics[width=0.7\columnwidth]{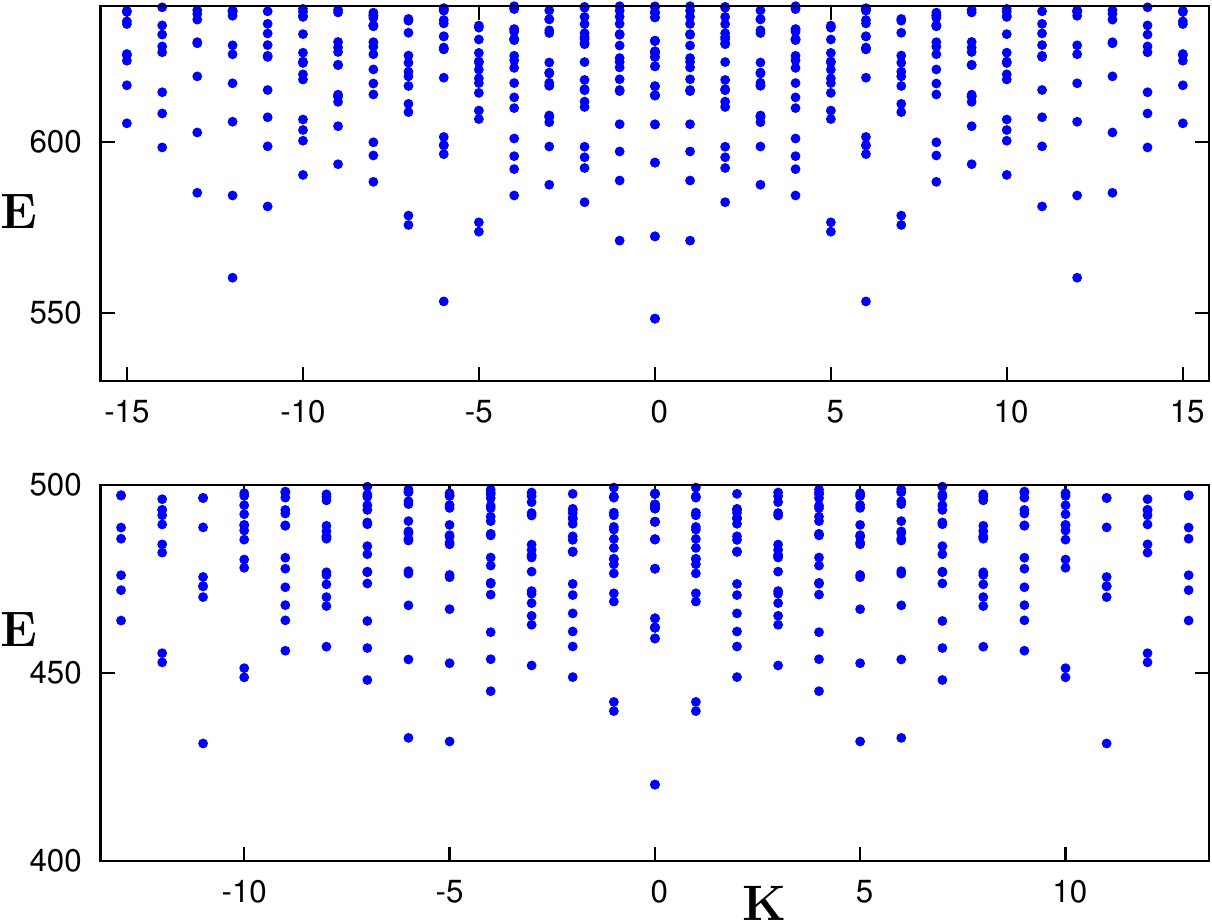}
\caption{Top panel~: the edge excitation sector for an even number of electrons $N_{e}=12$. The absolute ground state
is at $K=0$ and generates the identity plus energy tower of the Ising sector in addition to the charge excitations.
This structure of states repeats itself for $K=\pm N_e, \pm 2N_e, \dots$. This explains only half of the towers.
For $K=\pm N_e/2,\pm3N_e/2,\dots$ we assign the tower structure to the $\sigma$ tower of states with
the extra charge modes. This is consistent provided the velocities $v_n$ and $v_c$ are almost equal.
Bottom panel~: the odd number case with $N_e=11$. Now the absolute ground state is the twisted Pfaffian state
Eq.(\ref{oddTwisted})
and it supports the $\mu$ tower of states which is identical to the $\sigma$ tower. Next to it, we observe
the Majorana tower of states with charateristic (quasi) doubly degenerate ground states. Energies are in units of $\beta$.}
\label{pfspectrum}
\end{figure}

\begin{figure}[htb]
\includegraphics[width=0.7\columnwidth]{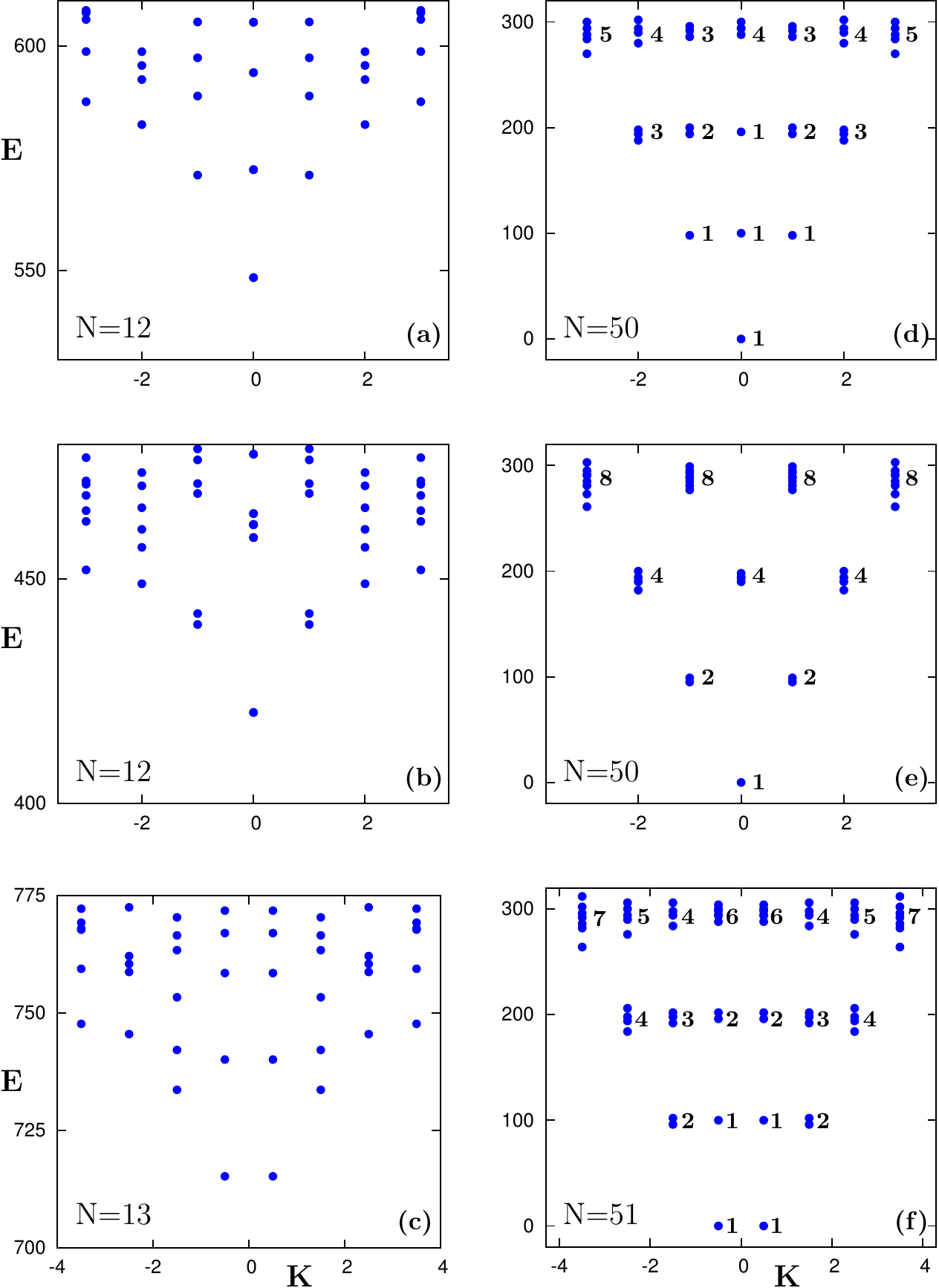}
\caption{Identification of CFT towers of states in the FQH regime where the cylinder
is large enough (left-hand side) and in the TT limit (right-hand side). 
The identity+energy tower is displayed in (a) and (d).
The $\sigma$ tower is in (b) and (e). Finally the Majorana tower
is given in (c) and (f).When the cylinder radius is sent to small values, the TT limit, the conformal towers become obvious.
In the TT limit the numbers are the degeneracies obtained in Table II.
Energies are in units of $\beta$.}
\label{TTtowers}
\end{figure}

We turn to the ED analysis in the cylinder geometry. The wavefunctions for the edge modes can be written easily
through the conformal transformation $z\rightarrow \exp (2i\pi z/L)$. As in the case of the primary Laughlin fluid,
we consider the case where there are extra orbitals available in the Fock space beyond those required by the fiducial flux-number of
particle relationship. We add a parabolic potential small enough so that there is no mixing between bulk and edge modes.
We find the typical low-lying spectrum displayed in Fig.(\ref{pfspectrum}). There are clear extremal states that are at the bottom of well-defined
towers of excited states. For $K_{tot}=\pm N, \pm 2N, \dots$ these states are global rigid translations of the ground state found at $K_{tot}=0$.
Strictly speaking this statement becomes exact only in the absence of $K$-space hard walls. Examination of the root
configuration of these states reveal their translated nature. They are generated by threading exactly one flux quantum
though the cylinder, according to the standard Laughlin gauge argument. However these states do not exhaust
the full set of extremal states. There are also arches terminating right in between at $\Delta K_{tot}=\pm N/2$.
These states can be generated by transferring a quasiparticle of charge $e/4$ from one side of the cylinder to the other side.
This leads to a change of sector in the sense of the previous section. Indeed the operation of quasiparticle transfer
is realized by the operator $\sigma {e}^{i\phi/2}$ in the CFT formulation. The $\sigma$ field changes the boundary
conditions of the fermion field while the vertex operator ${e}^{i\phi/2}$ takes care of the $e/4$ charge degree of freedom~\cite{Moore91,wen93}.
So we expect that between extremal states with NS-fermion
excitations there should be $\sigma$ sectors with R-fermion excitations. The conformal towers on top of these ground
states are different. For even number of particles we have the fully paired Pfaffian and its translated images. They support
bosonic excitations and the Ising tower involving the identity operator as well the energy operator~\cite{MalteBook}.
By a shift of $\Delta K_{tot}=N/2$ we find the tower of states associated with the spin field $\sigma$ and again the boson modes. This scheme
is repeated till the droplet is squeezed against the hard wall we impose in the Fock space. For odd number of particles,
the alternation is different because the Pfaffian now involves necessarily unpaired fermions. So we have towers generated
by the Majorana fermion field with NS boundary condition. Again by transfer of a $e/4$ quasiparticle we switch to the spin $\mu$ sector
which has the same structure as the $\sigma$ sector in the N even case. Our findings are in perfect agreement with the CFT scheme
for the low-lying states provided the Bose and Fermi velocities are equal. 
The identification of states becomes obvious in the TT limit as can be seen
in Fig.(\ref{TTtowers}). Here we plot the three distinct conformal towers on a wide cylinder in the left panel
and in the TT limit in the right panel. The CFT degeneracies are marked close to each multiplet of states.
The counting is in perfect agreement
with the CFT numbers obtained from Table (II).
In this TT limit it is easy to check that the Fermi and Bose velocities are indeed equal. This raises the question
whether this apparent equality on wide cylinders is a correct feature of the thermodynamic limit or if it is
simply a remnant of the TT limit. The ED data alone are not enough to answer convincingly this question
because the wide cylinders quickly require too many particles. We thus turn to another type of method
using the knowledge of wavefunctions which does not have this limitation.

\section{Measuring scaling dimensions by a Monte-Carlo method}
\label{MC}
While the conformal tower structure is readily apparent in ED studies even for a modest number of particles,
it is not easy to derive scaling dimensions within the confines of this method. In fact we find that only the global charge 
Luttinger parameter can be evaluated reliably. The knowledge of the Pfaffian CFT Ising$\times$U(1) predict also the scaling dimensions of all
conformal towers including notably the non-Abelian $\sigma$ tower with scaling dimension 1/8. We first note that treating the potential 
term in first-order perturbation theory is in fact enough
to describe the emergence of conformal towers by giving nonzero velocities to all low-energy modes. 
So another strategy is to use the exact knowledge
of  wavefunctions in the absence of the perturbing potential and to evaluate energies by taking the expectation value
of the potential with respect to these unperturbed states~: this is first-order perturbation theory with respect to the parabolic potential.
The full description of edge states given in section \ref{cylindersection} is adequate for
this purpose.

\begin{figure}[htb]
\includegraphics[width=0.9\columnwidth]{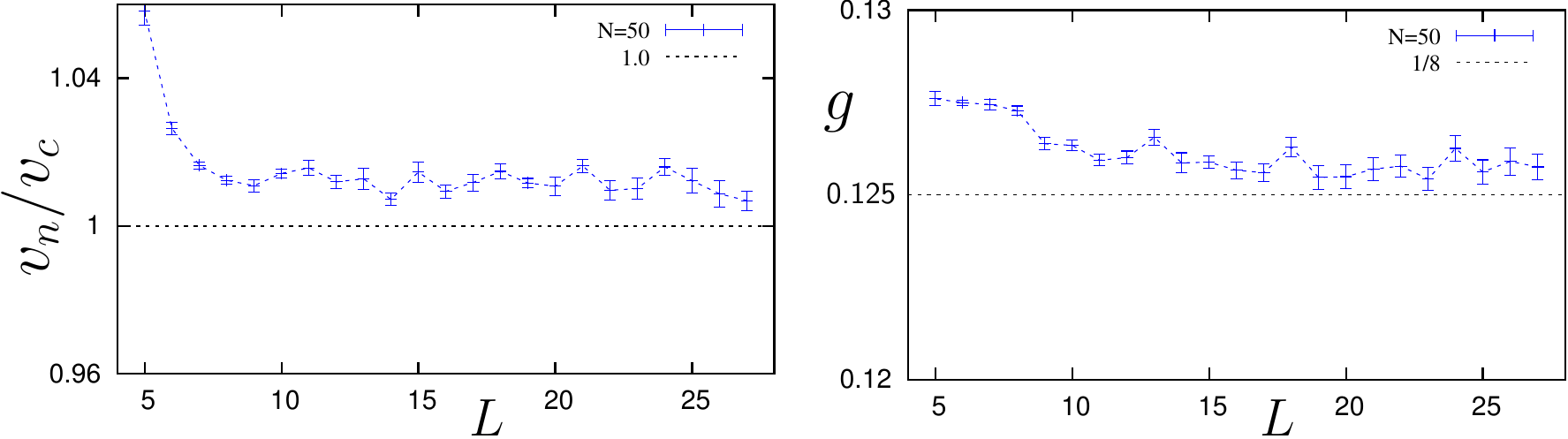}
\caption{Left panel~: The ratio of the Bose and Fermi velocities measured by the Monte-Carlo procedure
as a function of the perimeter of the cylinder $L$ for N=50 particles. The evidence is for the equality
of the two velocities in the thermodynamic limit as required by the interpretation of the conformal towers.
 We indicate confidence intervals  with a width equal to twice the standard
deviation. 
 Right panel : measurements of the Luttinger parameter of the bosonic part of edge modes using the same method. Results
are in agreement up to a few percent with the theoretical value $g=1/8$ at the thermodynamic limit ($N \rightarrow
+\infty$ and
$ L \rightarrow \infty$).}
\label{measures1}
\end{figure}
% \begin{figure}[htb]
% \includegraphics[width=0.4\columnwidth]{Figures/Fig_RayonMC.pdf}
% \caption{The charge boson radius (Luttinger parameter) as a function of $L$. It converges towards
% the expected 1/8 value.}
% \label{radius}
% \end{figure}
\begin{figure}[htb]
\includegraphics[width=0.4\columnwidth]{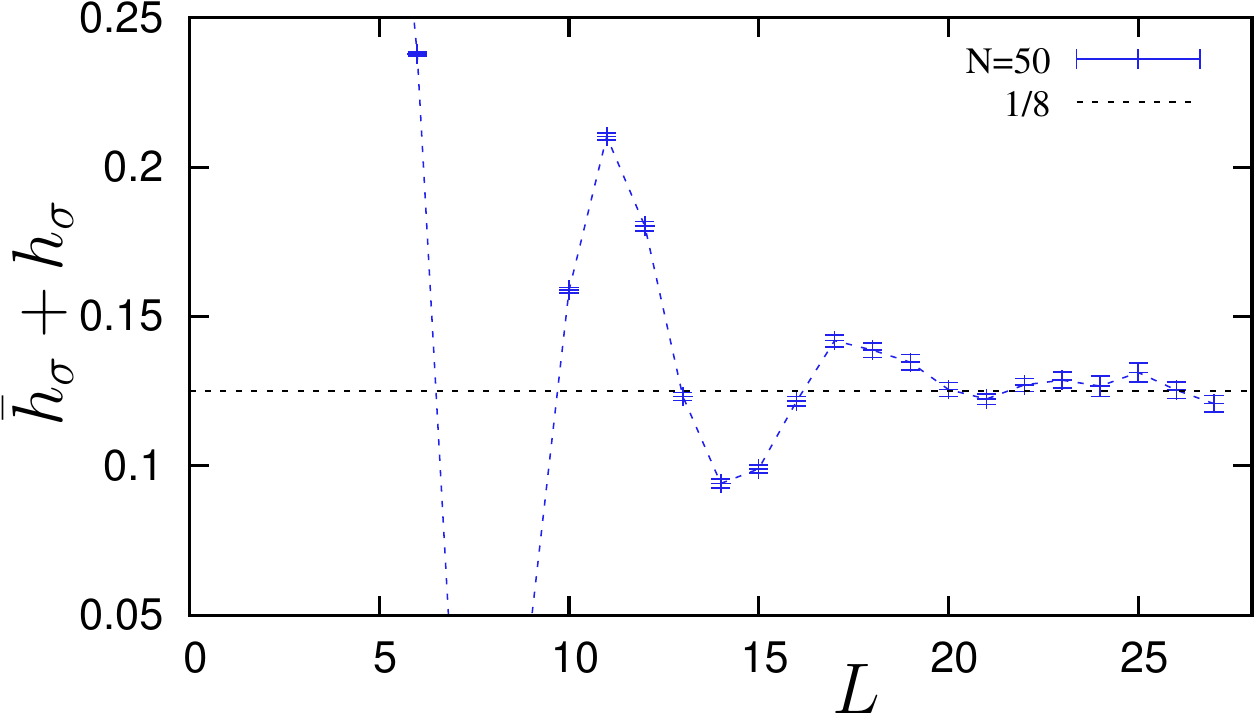}
\caption{The scaling dimension of the $\sigma$ spin field computed by the energy of lowest-lying state of
the corresponding conformal tower. It converges towards the expected value 1/8. Note that we need large values
of $L$ to obtain convergence (for N=50).}
\label{sigmadim}
\end{figure}

% From the many-body edge spectrum, we are able to estimate numerically the velocity of bosonic mode ($v_c$) and the
% velocity of the Majorana fermion mode ($v_n$). We are also able to check that the Luttinger parameter (or the radius
% of the bosonic field) ($g$), the conformal weight of the $\sigma$ sector ($\bar h_\sigma + h_\sigma)$ and
% the conformal weight of the Majorana sector ($\bar h_\theta + h_\theta)$ are in agreement with the CFT predictions:
% $g=1/8~,~~\bar h_\sigma + h_\sigma=1/8~,~~\bar h_\theta + h_\theta=1/2$ . 

We first consider the case of an even number of particles with the center
of mass of the Pfaffian wavefunction coinciding with the minimum of the parabolic confining potential.
The spectrum of edge modes for $N=12$ in such a configuration is shown in the top panel of Fig.(\ref{pfspectrum}). If we
note $E_n(K)$ the nth lowest energy at momentum $K$, one can estimate the
velocities $v_c$ and $v_n$ inside the identity+energy CFT tower on top of the Pfaffian state by the following differences~: 
\bea
\label{MesVelo1}
 E_1(1)-E_1(0) = \frac{2\pi}{L} v_c , \\
 E_2(0)-E_1(0) = \frac{2\pi}{L} v_n ,
\label{MesVelo2}
\eea
at leading order in the number of particles. Every conformal tower for fixed number of particles can be indexed by
the current quantum number $J=n_R-n_L$ (even) which counts the transfer of fractional charges $e n_{L,R} / 4 $
between edges, hence $J=0,\pm 2, \pm 4,\dots $. Ground states of those towers have momenta $K= J\, N/4 $, and their relative energies are given by
the Luttinger parameter $g$ which characterizes the energy needed to transfer a charge $e/4$ from one side to the
other of the droplet, and the scaling dimension of the spin primary field, $\bar h_\sigma + h_\sigma = 1/8$, which
gives the cost in energy to change the boundary conditions of the $\theta$ field. 
More precisely, if we look at  successive conformal towers indexed by the current quantum number $J$, the
ground state energies of each tower are, according to the effective theory explained in the previous section~: 
\be
E_1\left( \frac{1}{4} J\,N \right) - E_1(0) = \frac{\pi}{2 L} v_c g  J^2 + \delta_{J/2,odd}\, \frac{2 \pi}{L} v_n (\bar h_\sigma
+h_\sigma) ,
\label{MesDimSigma}
\ee 
where $\delta_{J/2,odd}$ is unity if $J/2$ is odd, zero otherwise.
In particular, we deduce the $g$ parameter from Eqs(\ref{tr1},\ref{tr2}) and the estimation of $v_c$ through:
\be
\label{MesG}
E_1(N)-E_1(0)=\beta N= \frac{2 \pi}{L}
v_c g\, .
\ee

We can use exact diagonalizations up to $N\le14$ to measure velocities $v_c$ and $v_n$ from 
Eqs.(\ref{MesVelo1},\ref{MesVelo2}). For better precision for this small number of particles, we also take into
account the next-to-leading
order in variation with $N$ in Eqs.(\ref{MesVelo1},\ref{MesVelo2}). At fixed $L$ and fixed value of the confining potential $\beta$, the
velocities $v_c$ and $v_n$ scale as $N$ because the slope of the parabolic confining potential at the
position of the edge grows like $N$ at leading order. We assume the simple scaling: $v=\alpha N + \gamma +
O(1/N)$ which is true in the TT limit. To measure the leading term, we first check that $N$ is large
enough such that $v_{c,n}(N)$ are linear, and then take the slope obtained with the last two points from the largest values of
$N$. 

We observe that $v_c \approx v_n$ for $L\lesssim15 $, i.e. when the two counterpropagating  edge are far
enough and there is no interaction between them. From the $v_c$ measurement, we are also able to compute the Luttinger
parameter $g$. We obtain $g\approx1/8$ as expected in the whole range $0\le L \lesssim15 $. However the practical limit
$N\le14$ does not allow to measure the conformal
dimensions $\bar h_\sigma + h_\sigma$ and $\bar h_\theta + h_\theta$. 

To circumvent the size limit, we use the Metropolis Monte Carlo
algorithm to compute energies of eigenstates for which we have an exact first quantized expression. Indeed, as long as
the external potential is infinitesimal, we know the explicit expressions  of some of the low-lying states. As
previously mentioned, the highest-density state which  lives at the
bottom of the identity+energy tower is the cylinder Moore-Read Pfaffian wavefunction deduced from $\Psi_{MR}$ of Eq.(\ref{pfaffianMR}). Then,
$E_1(0)=\langle \Psi_{MR} | \sum_i V(x_i) | \Psi_{MR} \rangle$, which can be computed directly in real space through
a 2N-dimensional integration. For $N=50$ and $O(10^9)$ Monte-Carlo steps, we obtain a relative precision
of  $10^{-5}$ over the energy. Other energies can be obtained in the same way. The energy $E_1(1)$ is the mean energy of
a bosonic
excitation of momentum $K_{tot}=+1$ localized at the right edge of the Moore-Read state~:
\be
b_1^{R\dag} \Psi_{MR}(Z_i)\propto \left\{\sum \limits_i Z_i \right\} ~ \Psi_{MR} (Z_i).
\ee
The state which corresponds to one pair of Majorana fermion excitations of momentum $\pm 1/2$ at each edge of the droplet can be
written as~:
\be
\Psi_{-1/2,+1/2}=\mathcal{A} \left( Z_1^{-1}
Z_2^0 \frac{1}{Z_3-Z_4}\dots\frac{1}{Z_{N-1}-Z_N} \right) \prod_{i<j}(Z_i-Z_j)^m ,
\ee
and its mean energy is $E_2(0)$. At the bottom of the $\sigma$ tower we have the twisted Pfaffian wavefunction~:
\be
\label{Twisted}
\Psi^{Twisted}=\mathrm{Pf}(\frac{Z_i+Z_j}{Z_i-Z_j})\prod_{i<j}(Z_i-Z_j)^m,
\ee
whose mean energy is $E_1(N/2)$. From the four energies, $E_1(0)$,$E_1(1)$,$E_2(0)$ and $E_1(N/2)$, we can deduce by
Eqs.(\ref{MesVelo1},\ref{MesVelo2},\ref{MesDimSigma},\ref{MesG}) estimates of the velocities $v_c$, $v_n$, 
the charge Luttinger parameter $g$,
and the scaling dimension $\bar h_\sigma + h_\sigma$. 

Our results for 50 particles and $0<L\lesssim 30 $ are shown in Figs.(\ref{measures1},\ref{sigmadim}).
Each energy is computed with $5\times 10^9$ MC steps. To estimate the standard deviation of each
point we use the binning method. Indeed, we separate the process into $M$ successive bins, average
energies onto each bin, obtain $M$ independent estimations of the quantity to compute, and estimate the standard
deviation from the $M$ binned values. We choose a bin size of a few $10^8$ steps so that our $M$ estimations are
statistically independent. 
We find that $v_c\approx v_n$ up to $2-3$ percent for $0<L\lesssim 25 $. The
imprecision comes both from the MC fluctuations and the next-to-leading order in variation with $N$ 
in Eqs.(\ref{MesVelo1},\ref{MesVelo2}). 
This strongly suggests that the two velocities are equal at the thermodynamic limit.
Similarly, we observe for
$0<L\lesssim 25 $ that the Luttinger parameter $g\approx1/8$ as expected up to a few percent.
Estimations of $\bar h_\sigma + h_\sigma$ are in agreement with the expected $1/8$ for large enough
value of $L$. 
The correct Ising CFT exponent that we obtain for the non-Abelian sector built upon the $\sigma$ field is an
important check of the CFT construction.
%In the Tao-Thouless limit, this value tends to $3/8$ and large values of N, unreachable with exact
%diagonalizations, are needed to be able to see the transition. 

To measure the conformal dimension $\bar h_\theta + h_\theta$ of the Majorana field operator, we compute the energies of states in the $N$ odd
sector. For N odd, there is an odd number of Majorana fermions in the untwisted sector. We then use the three
wavefunctions~:
\bea
\Psi_{+1/2}&=&\mathcal{A} \left( Z_1^0 \frac{1}{Z_2-Z_3}\dots\frac{1}{Z_{N-1}-Z_N} \right) \prod_{i<j}(Z_i-Z_j)^m \\
b_1^{R\dag} \Psi_{+1/2}&=& \left\{\sum \limits_i Z_i\right\} ~ \Psi_{+1/2} \\
\Psi^{odd,Twisted}&=& \mathcal{A} \left( Z_1^0 \frac{Z_2+Z_3}{Z_2-Z_3}\dots\frac{Z_{N-1}+Z_N}{Z_{N-1}-Z_N} \right)
\prod_{i<j}(Z_i-Z_j)^m 
\label{oddTwisted}
\eea
and compute their energies in order to get an estimate of $\bar h_\theta + h_\theta$. We find that $\bar h_\theta +
h_\theta=0.50(3)$ for $L=25, N=50$, which is in agreement with the CFT expectation  $1/2$.

%%%%%%%%%%%%%%%%%%%%%%%%%%%%%%%%%%%%%%%%%%%%%%%%%%%%%%%%%%%
\section{conclusion}
\label{conclude}

We have studied the edge modes of the filling fractions $\nu=1/3$ and $\nu=5/2$ in the QPC geometry by
using model wavefunctions of Laughlin and Moore-Read type.
These functions are generated numerically by using ED of special hard-core Hamiltonians. They are believed to capture
the physics of the more realistic Coulomb potential for 2DEGs. In the case of the $\nu=1/3$ QPC we have shown
that this geometry is suited to explore the electron-quasiparticle duality. It allows measurement of the Luttinger parameter
in the case of electron tunneling in a simple way.

In the Moore-Read Pfaffian case, we have studied in detail the conformal tower of states that are expected
from the 2D CFT foundations of this universality class of incompressible quantum fluids. The cylinder geometry
allows us to study the non-Abelian $\sigma$ sector in a simple way~: the corresponding CFT tower is found
by choosing the appropriate total momentum. There is no need to impose by hand the presence of a $e/4$
quasiparticle as is the case in the sphere or disk geometry. The CFT towers becomes very clear in the TT limit.
Extracting the correct scaling dimensions requires however to choose a regime with reasonably large
circumference $L$ and enough electrons. While ED is enough to get the global charge Luttinger exponent,
this does not allow the determination of the other scaling dimensions. For this purpose, we have introduce
a simple Monte-Carlo method based on the exact formulas for edge modes given in Ref.(\onlinecite{MR96}).
This Metropolis evaluation of energies in a perturbing potential leads to estimates of the scaling of the spin field
$\sigma$ as well as the Majorana field $\theta$ that are in agreement with the Ising CFT values.
We emphasize that this strategy is applicable to models whose relevant wavefunctions are given by an analytical
formula for which an efficient metropolis update is feasible.
While previous works~\cite{RJ,RR} have given evidence for the correct CFT counting
of non-Abelian quasiholes, we have in this paper uncovered the complete Virasoro structure of the CFT towers.
The equality of Bose and Fermi velocities of the edge modes also implies as noted in Ref.(\onlinecite{MR96})
that the special theory defined by the three-body hard-core model Eq.(\ref{H3}) has a hidden superconformal N=2 symmetry
at level k=2. We have found that a similar symmetry enhancement happens also for the bosonic Pfaffian at $\nu=1$
which has Kac-Moody symmetry SU(2) at level k=2.

Finally we stress that the potential added along the cylinder is by no means limited to the simple parabolic
shape we have explored. It may be modified to explore more complex, experimentally relevant configurations.
Recent work~\cite{Zaletel2013,Varjas2013} have also explored the cylinder geometry with the entanglement spectrum point of view
and offer a complementary view of the 2D CFT-based FQH wavefunctions.
% It is important to note
% that while the special CFT at the basis of the Moore-Read state is directly encoded in the structure of the wavefunction by construction
% it is
% not obviously encoded in the special three-body Hamiltonian Eq.(\ref{H3}) (although it is certainly plausible).

%%%%%%%%%%%%%%%%%%%%%%%%%%%%%%%%%%%%%%%%%%%%%%%%%%%%%%%%%%%
\begin{acknowledgments}
We acknowledge discussions with  P. Roche.
\end{acknowledgments}

%%%%%%%%%%%%%%%%%%%%%%%%%%%%%%%%%%%%%%%%%%%%%%%%%%%%%%%%%%%
%\bibliography{edges}
%%%%%%%%%%%%%%%%%%%%%%%%%%%%%%%%%%%%%%%%%%%%%%%%%%%%%%%%%%%

%%%%%%%%%%%%%%%%%%%%%%%%%%%%%%%%%%%%%%%%%%%%%%%%%%%%%%%%%%%%%%%%%%%%%%%
\end{document}